\documentclass[a4paper]{article}

\usepackage{makecell}
\usepackage{xcolor}
\usepackage{url}
\usepackage{nicefrac}
\usepackage{booktabs}
\usepackage{multirow}
\usepackage{floatrow}
\usepackage{INTERSPEECH2019}
\usepackage{amssymb}
\usepackage{pifont}

\usepackage[implicit=false]{hyperref}

\title{The xx205 System for the VoxCeleb Speaker Recognition Challenge 2020}
\name{Xu Xiang}
\address{AISpeech Ltd, China
\email{xu.xiang@aispeech.com}}

\begin{document}

\maketitle
\begin{abstract}
This report describes the systems submitted to the first and second tracks of the VoxCeleb Speaker Recognition Challenge (VoxSRC) 2020, which ranked second in both tracks.

Three key points of the system pipeline are explored: (1) investigating multiple CNN architectures including ResNet, Res2Net and dual path network (DPN) to extract the x-vectors, (2) using a composite angular margin softmax loss to train the speaker models, and (3) applying score normalization and system fusion to boost the performance.
Measured on the VoxSRC-20 Eval set, the best submitted systems achieve an EER of $3.808\%$ and a MinDCF of $0.1958$ in the close-condition track 1, and an EER of $3.798\%$ and a MinDCF of $0.1942$ in the open-condition track 2, respectively.

\end{abstract}
\vspace{10pt}


\section{Introduction}



The first and second tracks of the VoxSRC 2020 challenge allow participants to train the speaker model in a supervised manner.
In track 1, only the VoxCeleb2~\cite{Chung18a} development dataset that comprises 1,092,009 utterances from 5,994 speakers is allowed to train the model. While in track 2 any other data including that which is not publicly released except the challenge's test data can be used.
Different from the last year's challenge, there is an explicit domain shift between the training and test data. In addition, some utterances in the test data are much shorter than that of the training data.

The following sections describe the systems submitted to the challenge tracks 1 and 2 in detail.

\section{System}

\subsection{Data processing}
\label{sec:data}

\subsubsection{Data selection}
Table~\ref{tab:data_select} lists the dataset statistics used for training the speaker models. To comply with the requirements, only the development set of VoxCeleb2 corpus is used in track 1, whereas an additional Librispeech~\cite{panayotov2015librispeech} dataset is used in track 2.

\begin{table}[!htb]
    \caption{Training data for track 1 \& 2.}
    \centering
    \begin{tabular}{c|c|c|c} \toprule
         Track & Dataset & \#speakers &\# utterances \\ \midrule
         1& VoxCeleb2 Dev & 5994 & 1029009 \\\midrule
         \multirow{2}{*}{2}&  +Librispeech & 2338 & 281241 \\
         & Total &  8332 & 1310250
         \\\bottomrule
    \end{tabular}
    \label{tab:data_select}
\end{table}

\subsubsection{Data augmentation}
To increase the quantity and diversity of the training data, the data augmentation in the Kaldi~\cite{povey2011kaldi} VoxCeleb recipe\footnote{\url{https://github.com/kaldi-asr/kaldi/tree/master/egs/voxceleb}} is applied. As a result, 4 extra copies of the original training data are generated:
\begin{enumerate}
    \item The original audios
    \item Audios reverberated using RIRs~\cite{ko2017study}
    \item Audios augmented with MUSAN~\cite{snyder2015musan} noise
    \item Audios augmented with MUSAN music
    \item Audios augmented with MUSAN babel
\end{enumerate}

\subsubsection{Feature preparation}
40-dimensional Fbank is used as the acoustic feature in all the experiments, while no voice activity detection (VAD) module is involved. All the features are mean normalized with a sliding window of up to 3 seconds.

\subsection{Architecture}

\subsubsection{ResNet}
\begin{figure}[!th]
    \centering
    \begin{minipage}{0.5\textwidth}
    \centering
    \includegraphics[scale=0.6]{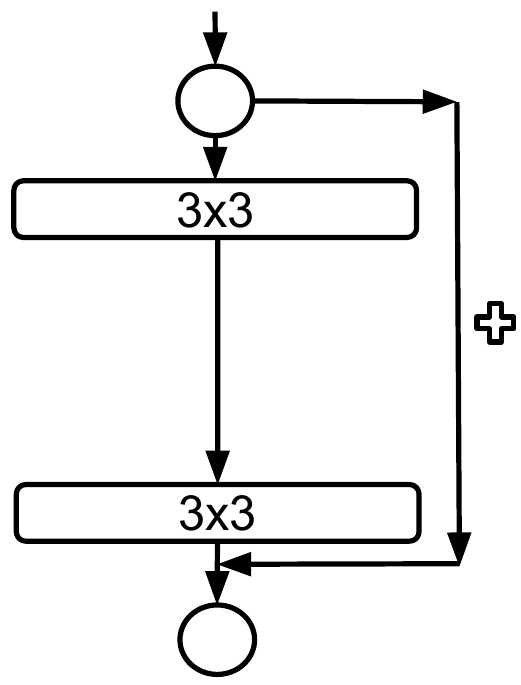}
    \end{minipage}\hfill
    \centering
    \begin{minipage}{0.5\textwidth}
    \centering
    \includegraphics[scale=0.6]{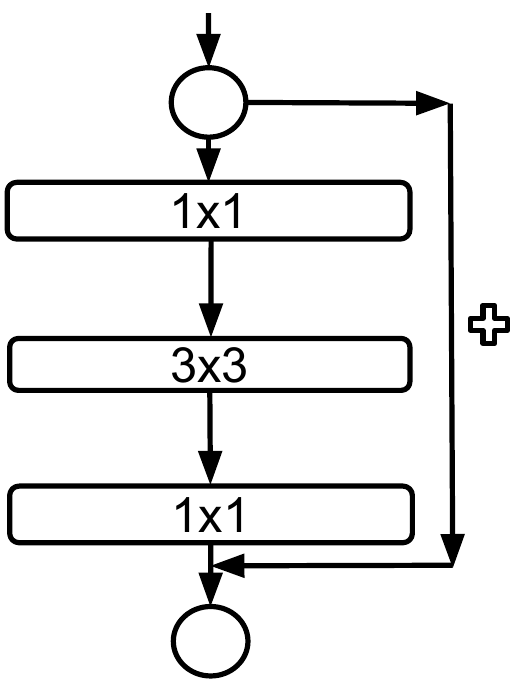}
    \end{minipage}
    \caption{Single basic ResNet block (left) and single bottleneck ResNet block (right).}
    \label{fig:res}
\end{figure}

ResNet has been successfully applied to speaker recognition~\cite{zeinali2019but} and is adopted here.
Table~\ref{table:model} shows the common used ResNet architecture for speaker modeling.
The input features is first processed by the initial convolution and the following 4 residual blocks, and then fed into the statistics pooling layer to aggregate the frame level information into an utterance level representation, and finally it is transformed into a fixed dimensional vector, or x-vector.
In this work, instead of the vanilla ResNet, two variants SE-ResNet and ResNeXt are used.
The residual block structure is depicted in Figure~\ref{fig:res}.

\begin{table}[h]
 \caption{ResNet architecture. {$\mathbf{L}$} is the number of the input frames, {\bf StatPool} denotes the statistics pooling.}
  \centering
  \label{table:model}
  \begin{tabular}{l c c c}
  
  \toprule

   Layer & Kernel size  & Stride& Output shape   \\
  \hline
  Conv 1 & $3 \times 3 \times 64$ & $1\times1$& $L \times 40\times 64$ \\
  \hline
  Blocks 1 & $3\times 3 \times 64$ & $1\times1$ & $ L \times40 \times 64$ \\
  \hline
  Blocks 2 & $3\times 3\times 128$ &$2\times2$ & $ \nicefrac{L}{2} \times20 \times 128$ \\
  \hline
  Blocks 3 & $3\times 3 \times 256$ &$2\times2$ & $ \nicefrac{L}{4} \times10 \times 256$\\
  \hline
  Blocks 4 & $3\times 3\times 512$ &$2\times2$ & $ \nicefrac{L}{8} \times5 \times 512$\\
  \hline
  StatPool & - & - & $5 \times 1024$ \\
  \hline
  Flatten & - & - & $5120$\\
  \hline
  Linear & - & - & $ 256 $ \\
  \bottomrule
  \end{tabular}
\end{table}

\subsubsection{Res2Net}
\begin{figure}[!th]
    \centering
    \includegraphics[scale=0.65]{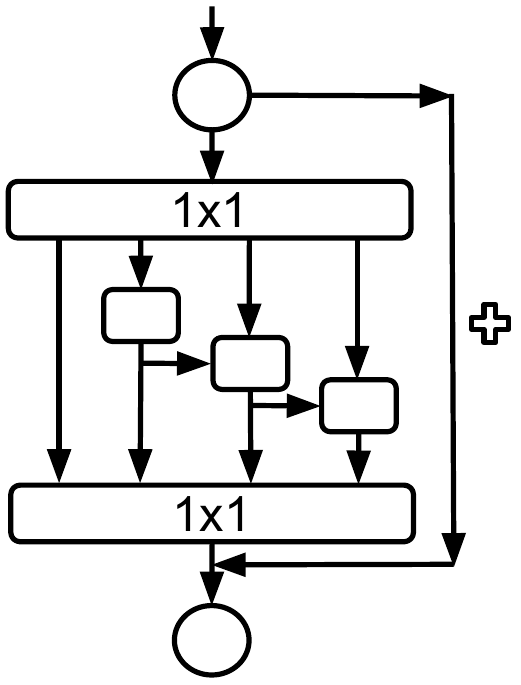}
    \caption{Single Res2Net block.}
    \label{fig:res2}
\end{figure}

Different from ResNet, Res2Net~\cite{gao2019res2net} enhances the bottleneck type residual block by introducing a more granular multi-scale structure, which is shown in Figure~\ref{fig:res2}.

\subsubsection{DPN}
\begin{figure}[h]
    \centering
    \begin{minipage}{0.5\textwidth}
    \centering
    \includegraphics[scale=0.6]{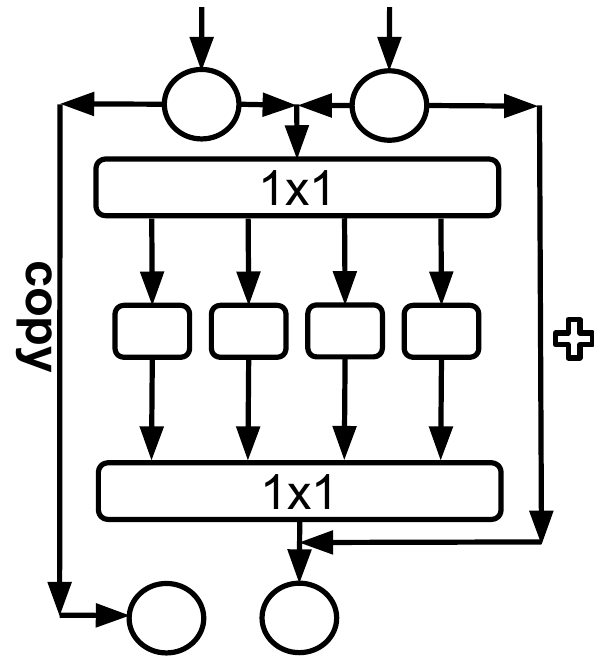}
    \end{minipage}\hfill
    \centering
    \begin{minipage}{0.5\textwidth}
    \centering
    \includegraphics[scale=0.6]{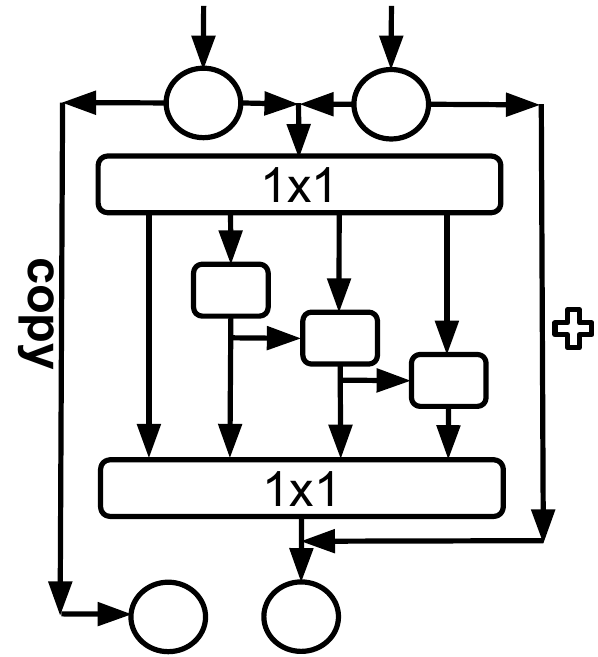}
    \end{minipage}
    \caption{Single DPN block (left) and single Res2DPN block (right).}
    \label{fig:dpn}
\end{figure}

DPN~\cite{chen2017dual} is an architecture inheriting both advantages of residual and densely connected paths, enabling effective feature re-usage and re-exploitation.
The dual path block structure (with optional Res2 enhancement) is illustrated in Figure~\ref{fig:dpn}.

\subsection{Loss function}

The additive margin softmax (AM-Softmax)~\cite{wang2018cosface} and additive angular margin softmax (AAM-Softmax)~\cite{deng2019arcface} loss functions that encourage inter-class separability and intra-class compactness have been successfully applied to speaker recognition~\cite{xiang2019margin}.
In this challenge, a composite margin softmax (CM-Softmax) loss function that takes the AM-softmax or AAM-softmax loss as a special case is proposed for training the speaker models,
$$L_\text{CM}=-\frac{1}{N}\sum_{i=1}^N\log \frac{e^{s({\cos(\theta_{{y_i}, i}+m_1)-m_2})}}{{e^{s({\cos(\theta_{{y_i}, i} + m_1)-m_2)}} + \sum_{j\neq i} e^{s\,{\cos(\theta_{{j}, i})}}}}$$
where $\theta_{j,i}$ is the angle between the class prototype vector (a column vector of the projection weight matrix) and
the input vector, $y_i$ is the ground truth class, $s$ is a scaling factor, and $m_1,m_2$ are two positive margins.

\subsection{Scoring and system fusion}
PLDA scoring and cosine scoring are used in the experiment.
The scores can be normalized by adaptive symmetric normalization (adaptive s-norm)~\cite{matejka2017analysis}.
For system fusion, the Bosaris toolkit~\cite{brummer2013bosaris} is used to train a linear fuser.

\section{Experiments}

\subsection{Dataset}

The models of the challenge track 1 are trained on the development set of VoxCeleb2.
The Librispeech corpus is also used for training in track 2. The VoxCeleb1 test sets~\cite{Nagrani17} and the VoxSRC 2020 validation set~\cite{chung2019voxsrc} are used to assess the single system performance. The performance of the fusion system is given on the VoxSRC 2020 evaluation set.

During training, the utterances are randomly cropped to 320-frame chunks\footnote{The frame length of 320 is a default setup, and more configurations can be found in Table~\ref{tab:exp_results}} for efficient fixed length training.
During testing, the utterances are not cropped.

\begin{table*}[h]
  \caption{The experimental results on the VoxCeleb1 test sets and the VoxSRC 2020 validation set. The \textbf{last two systems} are trained on the open-condition data. $^a$: apply SpecAugment~\cite{park2019specaugment} to the features during training, $^b$: use cosine decay instead of linear decay to adjust the learning rate, $^c$: the number of the input channels for each residual block is a multiple of 96 instead of 64, $^d$: use Multi-layer Feature Aggregation introduced in~\cite{desplanques2020ecapa}, $^e$: $2\times$ the batch size, $^f$: the number of the input frames is randomly chosen from 200 to 400.}
  \begin{tabular}{lccccccccc}
    \toprule
    \multirow{2}{*}{\textbf{Conf}} &
    \multirow{2}{*}{\textbf{Model}} &
    \multicolumn{2}{c}{\textbf{VoxCeleb1}} &
    \multicolumn{2}{c}{\textbf{VoxCeleb1-E}} & 
    \multicolumn{2}{c}{\textbf{VoxCeleb1-H}} &
    \multicolumn{2}{c}{\textbf{VoxSRC-20 Val}} \\ 
    \cmidrule(lr){3-4} \cmidrule(lr){5-6} \cmidrule(lr){7-8} \cmidrule(lr){9-10}
    & &
    \multicolumn{1}{c}{\textbf{EER(\%)}} & \multicolumn{1}{c}{\textbf{MinDCF}} &
    \multicolumn{1}{c}{\textbf{EER(\%)}} & \multicolumn{1}{c}{\textbf{MinDCF}} &
    \multicolumn{1}{c}{\textbf{EER(\%)}} & \multicolumn{1}{c}{\textbf{MinDCF}} &
    \multicolumn{1}{c}{\textbf{EER(\%)}} & \multicolumn{1}{c}{\textbf{MinDCF$_{0.05}$}}\\
    
    \midrule
    A+$\alpha$+1 & DPN68 & 0.9573 & 0.0903 & 0.9916 & 0.1102 & 1.738 & 0.1667 & 3.045 & 0.1526\\
    A+$\beta$+1 & DPN68 & 0.9467 & 0.1024 & 0.9775 & 0.1068 & 1.741 & 0.1643 & 3.006
 & 0.1532\\
    B+$\beta$+1 & DPN68 & 0.8244 & 0.0786 & 0.9254 & 0.0974 & 1.668 & 0.1616 & 2.876 & 0.1450\\
    C+$\beta$+1 & DPN68 & 0.8297 & 0.0974 & 0.9754 & 0.1055 & 1.754 & 0.1692 & 3.009 & 0.1542\\
    B+$\beta$+3 & DPN68 & 0.835 & \textbf{0.0638} & 0.9109 & \textbf{0.0959} & \textbf{1.637} & 0.1593 & 2.865 & \textbf{0.1422}\\
    A+$\alpha$+4 & DPN68 & 0.8829 & 0.0976 & 0.9892 & 0.1067 & 1.77 & 0.1634 & 2.996 & 0.1508\\
    B+$\beta$+2 & DPN68 & 0.8669 & 0.0872 & 0.9327 & 0.1011 & 1.673 & 0.1569 & 2.848 & 0.1445\\
    
    B+$\alpha$+2 & DPN68 & 0.8616 & 0.0811 & 0.922 & 0.0989 & 1.669 & 0.1538 & 2.882 & 0.1439\\
    B+$\alpha$+1$^a$ & DPN68 & 0.7978 & 0.0838 & 0.9413 & 0.1038 & 1.679 & 0.1563 & 2.853 & 0.1458\\
    B+$\alpha$+1 & DPN68 & 0.7818 & 0.0663 & 0.9361 & 0.1026 & 1.689 & 0.1620 & 2.89 & 0.1469\\
    B+$\alpha$+1$^b$ & DPN68 & 0.8457 & 0.0826 & 0.962 & 0.1044 & 1.726 & 0.1646 & 2.96 & 0.1530\\
    B+$\alpha$+5 & Res2Net50 & 0.8137 & 0.0940 & \textbf{0.8968} & 0.0967 & 1.693 & 0.1536 & 2.896 & 0.1473\\
    B+$\alpha$+6 & Res2Net50 & 0.851 & 0.1154 & 0.9441 & 0.1016 & 1.754 & 0.1636 & 2.95 & 0.1500\\
    B+$\alpha$+5$^c$ & Res2Net50 & 0.8616 & 0.0853 & 0.9254 & 0.1035 & 1.68 & 0.1647 & 2.88 & 0.1445\\
    B+$\alpha$+1 & SE-ResNet34 & 0.8457 & 0.0780 & 0.9875 & 0.1000 & 1.748 & 0.1631 & 3.029 & 0.1508\\
    B+$\alpha$+3 & ResNeXt50 & 0.8882 & 0.1104 & 1.025 & 0.1095 & 1.855 & 0.1678 & 3.125 & 0.1561\\
    B+$\alpha$+1 & DPN101 & 0.7925 & 0.0782 & 0.9113 & 0.0961 & 1.67 & 0.1558 & 2.822 & 0.1471\\
    B+$\alpha$+1$^d$ & DPN50 & 0.9839 & 0.0945 & 1.072 & 0.1109 & 1.935 & 0.1734 & 3.265 & 0.1627\\
    B+$\alpha$+1$^e$ & DPN50 & 0.835 & 0.1066 & 0.9482 & 0.1026 & 1.729 & 0.1569 & 2.931 & 0.1511\\
    B+$\alpha$+1$^f$ & DPN50 & 0.8457 & 0.0917 & 1.014 & 0.1109 & 1.787 & 0.1677 & 3.074 & 0.1599\\
    B+$\alpha$+1$^f$ & DPN68 & \textbf{0.7712} & 0.0774 & 0.9616 & 0.1032 & 1.668 & 0.1568 & 2.846 & 0.1447\\
    B+$\alpha$+1 & DPN74 & 0.7872 & 0.0725 & 0.9334 & 0.0997 & 1.655 & 0.1617 & \textbf{2.815} & 0.1464\\
    B+$\alpha$+1 & Res2DPN68 & 0.9361 & 0.0913 & 1.017 & 0.1082 & 1.813 & 0.1724 & 3.064 & 0.1566\\
    B+$\alpha$+1 & SE-DPN68 & 0.7978 & 0.0735 & 0.912 & 0.1028 & 1.677 & \textbf{0.1535} & 2.829 & 0.1449\\
    \hline\hline
    B+$\alpha$+1 & DPN50 & 0.8457 & 0.0799 & 0.9523 & 0.1008 & 1.669 & 0.1594 & 2.873 & 0.1467\\
    B+$\alpha$+1 & DPN68 & 0.8137 & 0.0801 & 0.8806 & 0.0959 & 1.591 & 0.1512 & 2.731 & 0.1427\\
    \bottomrule
  \end{tabular}
  \label{tab:exp_results}
\end{table*}
\begin{table*}[!h]
\caption{The configurations used for training speaker models, where A to C give three margin configurations, $\alpha$ and $\beta$ denote two setups for the embedding dimension and 1 to 7 describe different block settings. A complete system configuration is a combination of these three sets of sub-configurations. For example, A+$\alpha$+1 denotes a system which sets margins to (0.15,0.05), x-vector dimension to 256 and the block specification of ``base width=64''.}
  \centering
  \begin{tabular}{cccl}
    \toprule
    \textbf{Conf} & \textbf{margins} & \textbf{x-vector dimension} & \textbf{block specification} \\\midrule
    A & (0.15, 0.05) & &  \\
    B & (0.2, 0.1) & &  \\
    C & (0.3, 0.1) & &  \\\midrule
    $\alpha$ & & 256 &  \\
    $\beta$ & & 512 &  \\\midrule
    1 & & & base width=64  \\
    2 & & & base width=96 \\
    3 & & & base width=128  \\
    4 & & & base width=256 \\
    5 & & & scale=4, width=24 \\
    6 & & & scale=4, width=48 \\
    \bottomrule
  \end{tabular}
    \label{tab:conf}
\end{table*}

\subsection{Implementation details}

All implementations are based on the TensorFlow framework~\cite{abadi2016tensorflow} using 8 NVIDIA RTX 2080ti GPUs with the Momentum optimizer.
The models are trained for 23 epochs where one epoch is defined as a full pass through the dataset {\em by 8 GPUs}.
The learning rate is linearly increased to the maximum value 0.16 in the first 3 epochs and decayed exponentially in the final 10 epochs.
The tuple of the two margins $(m_1,m_2)$ stays $(0.0, 0.0)$ in the first 3 epochs and is linearly increased to $(0.2, 0.1)$ in the following 10 epochs and keeps the same in the final 10 epochs.
The scaling factor used in CM-Softmax is $32.0$.
The weight decay rate of 1e-3 is applied to the Momentum optimizer.

The submitted systems for track 1 consists of 19 DPN models, 3 Res2Net models and 2 ResNet models with various configurations. Other than these models, track 2 submission includes 2 DPN models trained on the pooled VoxCeleb2 development set and the Librispeech dataset.

\subsection{Scoring}
Both cosine and PLDA scoring are implemented in two tracks.
The cosine scores between two speech segments is simply computed with the corresponding length-normalized x-vectors.
The PLDA scores are derived with a PLDA model and the preprocessed x-vectors.
There are two datasets used to train two PLDA models respectively:
\begin{itemize}
    \item The original VoxCeleb2 development set
    \item The VoxCelebCat, a concatenated version of the VoxCeleb2 development set, built by concatenating the subsegments belonging to the same original video into an utterance
\end{itemize}
Following the common practice, the x-vectors are first centered using the dataset mean, then projected to 256-dimensional vectors with LDA and finally length-normalized.

For adaptive s-norm, the corresponding cohort set contains of the speaker-wise averages of the preprocessed x-vectors of the dataset (VoxCeleb2 or VoxCelebCat).
Top 400 cohorts are selected to compute the corresponding mean and standard deviation.
The cosine scores are normalized with the VoxCeleb2 dataset and the two sets of PLDA scores are normalized with the VoxCeleb2 and VoxCelebCat datasets respectively.
Thus, there are 6 sets of scores in total for each speaker model.

\subsection{Evaluation protocol}

Two performance metrics are presented in the experiment: (1) the Equal Error Rate (EER); and (2) the minimum detection cost of the function (minDCF).
The parameters $C_{miss}=1$, $C_{fa}=1$ and $P_{target}=0.05$ are used for the VoxSRC 2020 validation and evaluation set, while $P_{target}=0.01$ is set for reporting the minDCF for the VoxCeleb1 test sets.
In tracks 1 and 2 of this challenge, the minDCF is the primary metric for evaluating the system performance.

\subsection{Results}

Table~\ref{tab:exp_results} reports the single system performance on the VoxCeleb1 test sets and the VoxSRC 2020 validation set. Each single system is assessed by the cosine scores after applying adaptive s-norm.
The detailed configuration for each speaker model in Table~\ref{tab:exp_results} is given in Table~\ref{tab:conf}.
Table~\ref{tab:voxsrc} shows the performance of the fusion system on the VoxSRC 2020 evaluation set.

\begin{table}[h]
  \caption{Performance of the best submitted fusion systems. The submission for track 2 is the fusion of the all the systems shown in Table~\ref{tab:exp_results}, while the submission for track 1 excludes the last two systems.}
  \centering
  \begin{tabular}{ccccc}
    \toprule
    \multirow{2}{*}{\textbf{Track}} & \multicolumn{2}{c}{\textbf{VoxSRC-20 Val}} & \multicolumn{2}{c}{\textbf{\textbf{VoxSRC-20 Eval}}} \\ \cmidrule(lr){2-3} \cmidrule(lr){4-5}
    & EER(\%) & minDCF$_{0.05}$ & EER(\%) & minDCF$_{0.05}$ \\\midrule
    1 & 1.866 & 0.0982 & 3.808 & 0.1958  \\
    \hline\hline
    2 & 1.818 & 0.0979 & 3.798 & 0.1942  \\
    \bottomrule
  \end{tabular}
    \label{tab:voxsrc}
\end{table}

For the close-condition track 1, the best performing single system is based on DPN68 (Conf: B+$\beta$+3), achieveing an EER of $2.865\%$ and minDCF of $0.1422$ on VoxSRC-20 Val. With score fusion, the fusion system achieves an EER of $1.866\%$ and a minDCF of $0.0982$ on VoxSRC-20 Val and an EER of $3.808\%$ and a minDCF of $0.1958$ on VoxSRC-20 Eval.

For the open-condition track 2, DPN68 based system still performs the best, obtaining an EER of $2.731\%$ and a minDCF of $0.1427$. The fusion system achieves an EER of $1.818\%$ EER and a minDCF of $0.0979$ on VoxSRC-20 Val and an EER of $3.798\%$ and a minDCF of $0.1942$ on VoxSRC-20 Eval.

\section{Acknowledgement}
The author would like to thank the support of AISpeech Ltd and Dr. Shuai Wang of SpeechLab, Shanghai Jiao Tong University.

\section{Conclusion}

The report describes the proposed systems for the VoxSRC Speaker Recognition Challenge 2020 in tracks 1 and 2. The speaker models are based on deep CNN architectures (ResNet, ResNeXt, Res2Net and DPN) and optimized using margin based loss function. In track 1, our best submitted fusion system achieves an EER of $3.808\%$ and a minDCF of $0.1958$. With additional training data, in track 2, our best submitted system achieves an EER of $3.798\%$ and a minDCF of $0.1942$.


\newpage
\raggedbottom
\bibliographystyle{IEEEtran}
\bibliography{main.bbl}
\end{document}